\documentclass[12pt]{article}
\begin{document} 

\begin{center} 

\Large{\bf  Large two-pion-exchange contributions to \\ 
the $pp\to pp\pi^0$ reaction }

\large{F. Myhrer }\footnote{E-mail: myhrer@physics.sc.edu}

\vspace{3mm}

{\it Department of Physics and Astronomy \\
University of South Carolina,  
Columbia, SC 29208 }
\end{center}

\begin{abstract}
The large two-pion exchange amplitudes are 
calculated in HB$\chi$PT and their net 
contribution to the reaction cross section is large. 
\end{abstract}


\vspace{5mm}

Refs.~\cite{pmmmk96,cfmv96}
evaluated this reaction at threshold in HB$\chi$PT. 
They found that the impulse approximation (I.A.) 
and one-pion-exchange (Resc) diagrams 
interfere 
destructively resulting in a 
very small reaction cross section. 
According to Lee and Riska \cite{lr93} 
a model explanation of the measured reaction 
cross section 
near threshold 
requires the contributions from 
heavy meson ($\sigma$ and $\omega$) exchange 
in addition to the one-pion-exchange.  
%
The 
$\sigma$-meson-exchange 
is more properly described by correlated 
two-pion-exchange. 
This knowledge prompted a HB$\chi$PT study of 
two-pion-exchange (TPE) contributions to  
the $pp\to pp\pi^0$ reaction amplitude~\cite{dkms99}.

\vspace{3mm} 

In Weinberg chiral counting the 
TPE contributions 
are of higher chiral order in HB$\chi$PT. 
However, 
it was shown by Ref.~\cite{dkms99} 
that some TPE amplitudes are 
as large or larger than the lower chiral order 
Resc contribution. 
At threshold
the typical momentum is 
$p \sim \sqrt{m_\pi m_N}$. 
%
This large momentum 
prompted Cohen 
{\it et al.}~\cite{cfmv96} to propose a 
momentum counting rule, reviewed in 
Ref.~\cite{hanhart04}. 
According to this counting, 
one finds that 
the Resc diagram is higher order in $p/\Lambda$ 
compared to some ``dominant" TPE diagrams, and  
this counting agrees with the 
numerical evaluations of the TPE 
diagrams of Ref.~\cite{dkms99}. 
One drawback with the momentum counting is that 
the sum of the diagrams 
in each ``momentum order"
no longer is independent of the 
definition of the pion field. 
Hanhart and Kaiser (HK) \cite{hk02} 
used momentum counting to 
evaluate the ``leading" momentum behavior of 
the dominant TPE diagrams. HK also found that 
diagram II in Ref.~\cite{dkms99} should 
have opposite sign (which we confirmed),  
and they  
found that the sum of the 
leading momentum behavior of the 
three dominant TPE diagrams cancel.  
%
We will present  
results from Ref.~\cite{dkms99} and a 
recent calculations~\cite{kkms06} which show 
that the ``sub-leading" 
parts of a dominant TPE diagram gives a 
contribution comparable to  
the Resc amplitude. 

\vspace{3mm} 

The TPE 
transition operators (TO) 
were evaluated analytically by  
Ref.~\cite{dkms99} in HB$\chi$PT. 
When these operators are sandwiched between  
phenomenologically determined 
distorted 
$NN$ wave 
functions, 
the momentum integrals converge 
slowly~\cite{kkms06}. 
This slow convergence can be understood 
when we adopt the  threshold 
fixed kinematics approximation (FKA). 
Imposing 
FKA on the analytic expressions for the TO
given in  
Ref.~\cite{dkms99}, we make an 
asymptotic expansion in the two-nucleon 
momentum transfer 
($ |\vec{k} | 
= |\vec{p}-\vec{p}^{\; \prime}| 
\to \infty $). 
The TO matrix $T$ of 
the TPE diagrams 
is of the form 
\begin{eqnarray*}
T &=& \left( \frac{g_A}{f_\pi} \right) 
\left( \vec{\Sigma}\cdot \vec{k} \right) 
t(p,p^\prime , x)
\end{eqnarray*} 
where 
$x=\hat{p}\cdot\hat{p}^\prime$. 
The {\it asymptotic} momentum behavior for 
$ t(p,p^\prime , x)$ is 
$
t(p,p^\prime , x) \sim  
t_1 \; | \vec{k} | + t_2 \; 
{\rm ln}[\Lambda^2/ |\vec{k} |^2 ] 
+ t_3 + \delta t(p,p^\prime , x) , 
$
where $\delta t(p,p^\prime , x) $ is  
${\cal O}(|\vec{k} |^{-1}) $, 
and the amplitudes  
$t_i$, $i=1, 2, 3$ are known analytic 
expressions for each diagram. 
The $t_1$ amplitude is the dominant 
TPE amplitude of HK~\cite{hk02}. 
%
%
$$
\begin{array}{|l| |r| r|r|r|r|r|r|}
\hline 
{\rm Amplitude \;  K=} & I&II&III&IV&V&VI&VII \\ 
\hline 
 R_K    & -.70& -6.70& -6.70 & 9.50 &0.18&0.14&2.65 \\
\hline 
 t_1 \; \; \; \; \; \propto & - & -2  & -1  
& +3 &- &- & - \\
\hline 
\end{array} 
$$
In the table the row marked $R_K$, gives  
the values of the ratio of TO 
to the Resc amplitude in 
the plane wave approximation for 
the seven amplitudes of Ref.~\cite{dkms99}.  
%
As indicated in the last row of the table, 
marked $t_1$, 
the leading momentum terms of the 
TO from diagrams 
II, III and IV sum to zero,   
confirming HK's result~\cite{hk02}. 
The non-cancellation of the dominant amplitudes 
can however be inferred from the $R_K$ 
row since the ratios 
II:III:IV are not -2:-1:3 
but roughly -2:-2:3. 
When we  \underline{remove}  $t_1$,  
we find that 
the sum of the 
two-pion-exchange amplitudes 
is larger than 
the Resc amplitude \cite{kkms06}. 

\vspace{3mm} 

This research is supported in part by a 
grant from NSF.

\end{document}